\documentclass[a3paper,twocolumn]{article}
\usepackage{eurosis}
\usepackage{url}
\usepackage[sectionbib]{natbib}
\setcitestyle{notesep={; },round,aysep={},yysep={;}}
\usepackage{graphicx}
\usepackage{listings}
\usepackage[dvipsnames,table]{xcolor}

\begin{document}

\title{
	\vspace*{-2cm}
	{\normalfont This is a preliminary (and longer) version of the paper: Tricas, F. `A PROPOSAL FOR FEDERATED CHATBOTS FOR DISTRIBUTED INFORMATION ACCESS' submitted (and accepted) at the ESM®'2023 (The 37th annual European Simulation and Modelling Conference) October 2023, Toulouse, France, pp. 151--154.}

	\vspace*{2cm}
	A PROPOSAL FOR FEDERATED CHATBOTS FOR DISTRIBUTED INFORMATION ACCESS}
%
%
\author{Fernando Tricas García\\
%
%
Instituto de Investigación en Ingeniería de Aragón.\\
Universidad de Zaragoza, María de Luna, 1. Zaragoza, Spain.\\
E-mail: ftricas@unizar.es
}

{\let\newpage\relax\maketitle}

\date{}
\maketitle              

\begin{abstract}
	Chatbots can be a good way to interact with IoT devices, and other information systems: they can provide information with a convenient interface for casual or frequent interaction.
	Sometimes there can be good reasons to have more than one chatbot: maybe we have several computers, or diverse infrastructure, with different access conditions.
	This work concentrates on this case, when it can be useful to establish a method for them to work in a cooperative way.
In principle, coordination is a good property: each one of these
chatbots can be devoted to solve different tasks and our users can have different needs
when accessing to every capability of each chatbot.

In this paper we are proposing an architecture for several chatbots that can interact via a command and control channel, requesting actions for other bots and collecting the replies in order to pass them to the user.
The chatbot infrastructure is lightweight, and it can use public (but not
publicly viewable) infrastructure providing an easy way to start a project
with it.

\end{abstract}
\section{INTRODUCTION}

Chatbots can be easily used to get information from different sources:
IoT sensors, information from databases, results of the execution of
commands in a given infrastructure,...
Part of this infrastructure can be well connected
computers with direct access to the internet, but other part can be based on
machines with reduced capabilities, or even some not so powerful
devices to get information from different sources.

Moreover, maybe our users need to have access to the information in several ways: a desktop computer, some mobile device\ldots

Chatbots have been used  for knowledge management in maintenance processes~\citep{coli2020automatic}, in business process automation~\citep{rizk2020unified}, or in smart home contexts~\citep{8448761,5766475}.
We are not aware of federated chatbot proposals in the literature for this kind of work.

When talking about cooperative robots, the term Swarm robotics~\citep{Brambilla2013} comes to mind. Notice that in that case the cooperation is typically done in a physical sense, with local interactions among agents and between the agents and the environment.

Command and control techniques (C\&C) have been seen frequently in the cybersecurity literature, because it seems to be a simple (and effective) way of managing a botnet sending commands to the bots and, sometimes, reading the replies.
See for example,~\citep{10.1007/978-3-642-13708-2_30} and references.
Anyway, we are not aware of attacks so sophisticated to have a more complex coordination of the bots. The usual setup is a client-server configuration where the bots just check for new commands, updates, and so on.
They have used typically IRC communication protocol, but more modern approaches can used other mechanisms, such as: Simple Mail Transfer Protocol (SMTP), or even Twitter and other services.


There are some projects related to the access to information using chatbots, such as: Almond~\citep{Lam18}, which is a  virtual assistant that lets users share their digital assets easily in natural language, or
Jarvis~\citep{jarvis}, which is

\begin{quote}
... a simple personal assistant for Linux, MacOS and Windows which works on the command line. He can talk to you if you enable his voice. He can tell you the weather, he can find restaurants and other places near you.
\end{quote}

Chatbots are being used extensively to add another channel of communication between companies and customers~\citep{XuLiuGuoSinhaAkkiraju17} and most of the research is concentrated on the intelligence part, natural language processing (NPL), and so on~\citep{8592630}.

As previously stated we are proposing an architecture to explore the use of chatbots as an interface to some information, using different communication channels and different computers hosting the bots.

\section{THE ARCHITECTURE}
\label{secArch}

\begin{figure}[htbp]
	\centerline{\includegraphics[width=\columnwidth]{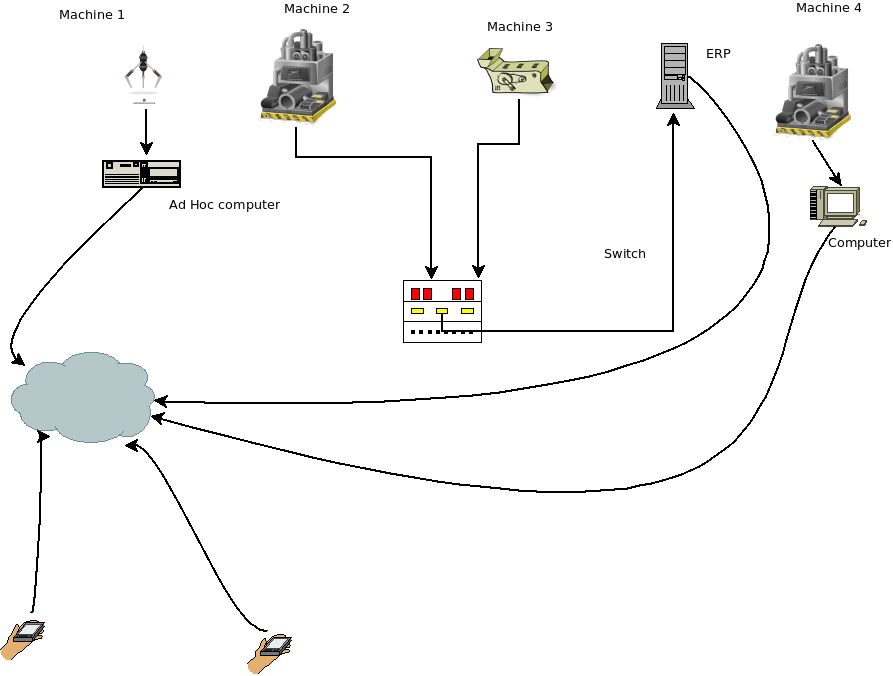}}
	\caption{Architecture of the system with several machines (Machine 1, 2, 3 4, some computers (Ad Hoc Computer, Computer and ERP computer,...) \label{figArchitecture}}
\end{figure}

We are proposing a system where the installation of the chatbots can be done in several computers, connected to  different communication channels (such as IRC, Slack, Discord, Gitter, Telegram, ...).
In this way, we will have a set of chatbots that can communicate with the users
in their respective channels.
They are listening (and replying) in different networks, maybe with diverse access control policies (some in public networks, some in private ones).

This architecture is sketched in Figure~\ref{figArchitecture}.
There we can see some machines attached to ad hoc computers (Machine 1, Machine 4), and others (Machine 2 and Machine 3) attached to the communication network of the company, and reporting to the ERP.
All of them can communicate with some chatbot, which is reachable from the internet using handheld devices.
Anybody (with adequate authorisation) could access each of these chatbots and obtain the information about its attached system.

The next step is to be able to pass information among them with two objectives:

\begin{itemize}
\item to add and manage as many chatbots reporting information as needed in a convenient way.
\item to provide capabilities to the user, who can ignore all of these settings and concentrate on requesting
and reading the information using her preferred way of communication (among the available ones).
\end{itemize}

To reach this we need to adapt our chatbots to interact with the others.
We have decided to use a common communication channel (the command and control channel) for this: each one of the bots can read and write there.

When a bot receives a request from some user it will check if it can fulfil the command:

\begin{itemize}
 \item{If it can, it will:
	\begin{enumerate}
	\item Execute the action,
	\item Write a message to the user with the reply.
	\end{enumerate}}
 \item{if not, it will post the command to the common channel and it will
wait to see if there is a reply.

	The other bots will read this command and check whether they can execute it or not, basically replying to these two questions:

	\begin{itemize}
		\item Do I know this command?
		\item{Can I execute this command with these parameters?

				Hopefully, one of them will be able to execute it and, when done, it will have some result to offer. If this is true, it will:


	\begin{enumerate}
		\item Delete the message requesting the action from the channel.
		\item Execute the action.
		\item Write a new message in the common channel which embeds
			the reply.

	\end{enumerate}

	The deletion of the message avoids some work to the other bots that have not read it (and to avoid duplicating the work). There can be cases where more than one must execute the command (and provide a reply); we could prepare special commands (multi-bot) for these cases.

}
\end{itemize}
	Each one of the other bots, will be able to read (while available) the reply and will
	check if:
	\begin{itemize}
		\item{Is this reply for me?

	Hopefully, one of them will be the originator of the command. It will
then:

	\begin{enumerate}
		\item Delete the message with the reply.
		\item Write a message to the requesting user with the information.
	\end{enumerate}

}
	\end{itemize}
}
\end{itemize}

\lstset{
  language=Python,
  basicstyle=\small\ttfamily,
  columns=fullflexible,
  breaklines=true,
  postbreak=\mbox{\textcolor{black}{$\hookrightarrow$}\space},
  showstringspaces=false,
  keywordstyle=\bfseries,
  commentstyle=\itshape\color{gray},
  stringstyle=\color{black}
}

Let us write a pseudo-program in Python for the shake of clarity, in Listing~\ref{listing-command}.

\begin{lstlisting}[label={listing-command}, caption={Bot Command Handling}]
def manage_command_in_channel(chan, msg):
    delete_request_message(chan, msg)
    reply = execute_action(msg)
    post_to_channel(chan, reply)

def manage_reply_in_channel(chan, msg):
    delete_reply_message(chan, msg)
    send_reply_to_user(msg)

def handle_msgs(chan):
    msgs = read_commands(chan)
    for msg in msgs:
        if (msg_is_a_command(msg) and command_is_for_me(msg)):
            manage_command_in_channel(chan, msg)
        if (msg_is_a_reply(msg) and reply_is_for_me(msg)):
            manage_reply_in_channel(chan, msg):

def forward_command(chan, command):
    msg = prepare_message(command)
    post_to_channel(chan, msg)

def handle_command(command):
    if bot_can_fulfill(command):
        msg = execute_command(command)
        send_reply_to_user(msg)
    else:
        forward_command(chan, command)


\end{lstlisting}

%
%

Notice that all of this can be very fast (if every bot is available and ready to reply) but it can also be slow: some bots can be offline, or only available part of the time...
The common channel guarantees that they will be eventually able to read the commands and replies available for them.


\section{A PROTOTYPE FOR EVALUATION\label{secPrototype}}

We have used an existing Chatbot project, Errbot~\citep{errbot}, written in Python and available under the GPL-3 license.
It is a non intelligent chatbot, so we must write the commands using some syntax conventions. It would be easy to add some NPL module if some intelligent behaviour is expected, but this is not the objective of this work.
The project provides a way to do fast prototyping for our problem by means of its extensive plugin framework. It allows us to concentrate on developing just the part of obtaining the information of our sensors. It has an interface with several available backends (Telegram, Slack, Discord, Gitter, IRC, ...) and a simple but adequate built-in administration with some security features.

In order to evaluate the proposal we have set up a simulated (but using
actual computers and real world actions and sensors) in a home/laboratory
environment.
For this, we have installed our chatbots in two Raspberry Pi
computers: one of them is the Raspberry Pi 4 Model B 
Processor rev 3 (v7l),
and the other is an older Raspberry Pi Model B Rev 2. We have also several old and modern desktop computers. 

\begin{figure}[htbp]
	\centerline{\includegraphics[width=\columnwidth]{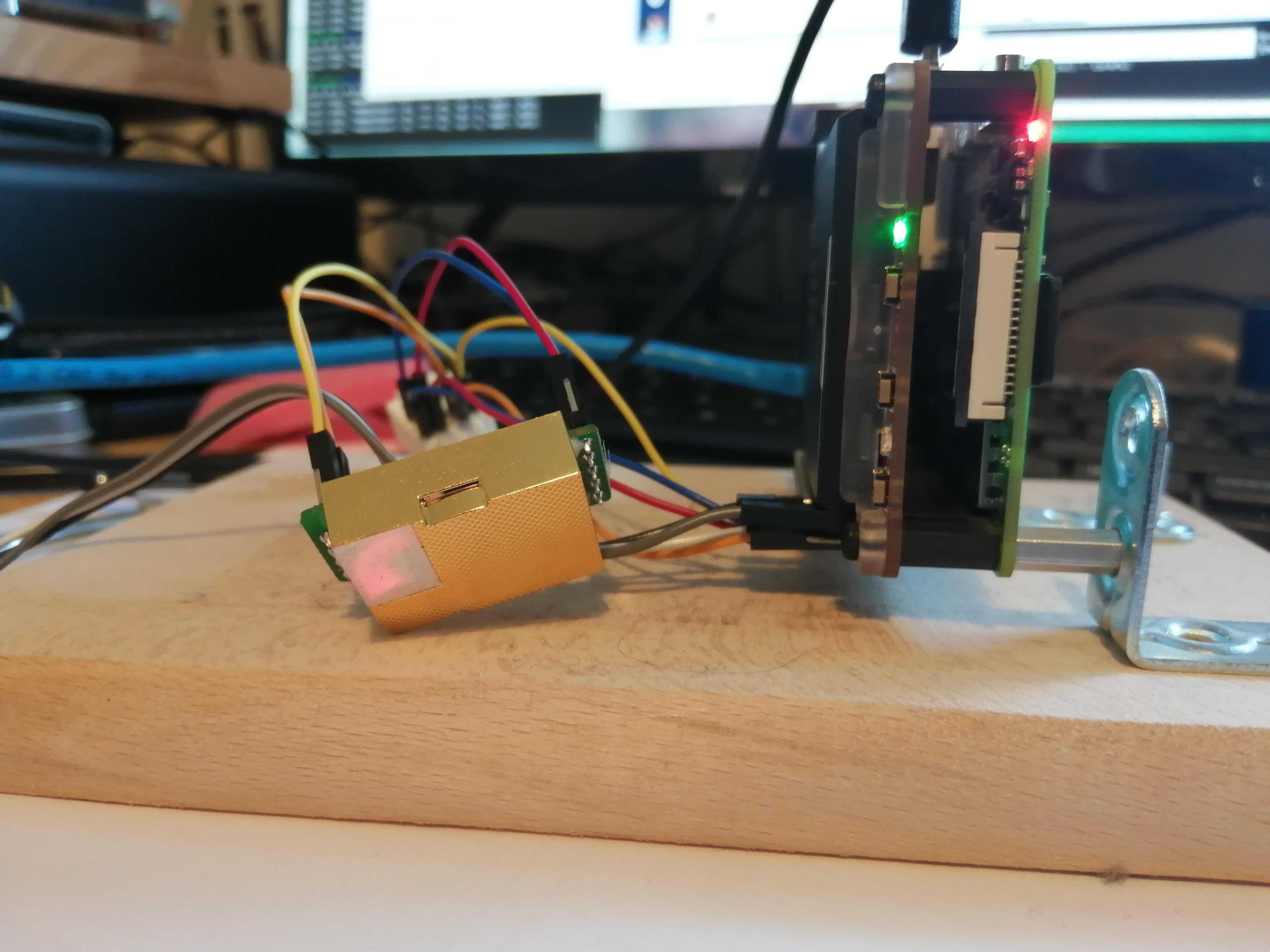}}
	\caption{A CO2 sensor attached to a Raspberry Pi computer \label{figSensor}}
\end{figure}

We have attached  several sensors, webcams, and other devices to these computers.
The sensors are cheap standard versions such as the MH-Z19C IR Infrared CO2 Sensor~\citep{doi:10.1021/acs.estlett.1c00183} (there is a picture of it in Figure~\ref{figSensor}), and the DHT22 Digital Temperature Humidity. The cameras are standard webcams attached to some PC, and the relay is the Sonoff Basic WiFi Smart Switch.

In this way, we can have some chatbots than can provide environment data (via sensors or images with the webcams), and one of then which can switch on an off some external device.

Everything is connected to some local network (which give each device read and
	write access to the internet; some of them are behind a firewall, so
	they are not publicly accessible from the outside), and they are in
	different geographically distributed locations.
Notice that we are trying to provide access to some `internal' information without
reducing the available in place access controls.


\subsection{Errbot Installation}

The installation of the Errbot software is easy to perform with the standard Python package manager, `pip'. In some cases, some more installation steps will be needed for the backend.
Each bot must have a backend interface installed and configured, which will
provide the way for the bot to communicate with users in some established
channel. These configuration steps include also the need to configure access to
this backed communication channel that can involve some credentials creation
and manipulation (name --nick--, and credentials).

Errbot has adopted a very simplistic way of managing credentials, via the inclusion of them in plain text in the configuration file.

For example, for the Slack backend configuration, we can setup the bot identity with:

\begin{quote}
\lstinline+BOT_IDENTITY = {'token':'xoxb-4326149011-a...', }+
\end{quote}


and the bot admins with:

\begin{quote}
\lstinline+BOT_ADMINS = ('@gbin', '@zoni', )+
\end{quote}


The configuration parameters are different for different backends.
Being the config file a Python file ({\tt config.py}) which is included in some moment at the chatbot start, it is easy to use any other approach, such as integrating the secrets storage in some database or, at least, in more robust systems of storage.

\subsection{Plugin Development}

Plugins are a convenient way of extending our chatbot, and they can be loaded from a configurable local directory or by means of the {\tt repos} command. This can be a convenient way to install, uninstall, upgrade and other tasks from a {\tt git} repository but we won't detail these options here.

The definition of a plugin has two parts, as stated in the documentation,
\begin{quote}
	\ldots a special file (.plug) and one or more Python (.py) files containing the actual code of your plugin ....
\end{quote}

We won't comment here on the first file, which is used in order to supply some meta-data to the chatbot (name, Python version, ...).

A plugin can be used to define one or more commands. We have included in Listing~\ref{listing-plugin} a very basic plugin, {\tt HelloWorld} (it will be the name of the class), with just one {\tt hello} command (each command is defined by a method).
There we can see that we need to use {\tt @botcmd} decorator in order to define a bot command. Without this directive the function (or method) will be considered internal.
	The method definition takes two parameters, {\tt msg} and {\tt args}. The first  represents the full message object received by Errbot (sender, channel, and other information). The second is a string (a list is also possible) with the arguments passed to the command. This characteristic will allow us, for example, to ask for different sensors or other commands if available.

\begin{lstlisting}[label={listing-plugin}, caption={Hello World plugin}]
from errbot import BotPlugin, botcmd

class HelloWorld(BotPlugin):
    """Example 'Hello, world!' plugin for Errbot"""

    @botcmd
    def hello(self, msg, args):
        """Say hello to the world"""
        return "Hello, world!"
\end{lstlisting}

	For example, the command could be used alone in the case we have just one temperature sensor (or in the case of requesting information from the default one):

\begin{quote}
\lstinline+temp+
\end{quote}


If we needed to select some sensor, we could use something like (see Subsection~\ref{subSecExamples} for more details):

\begin{quote}
\lstinline+temp room23+
\end{quote}


Here, \lstinline+room23+ is the parameter for the \lstinline+temp+ command.

	It also has logging capabilities, as expected in a project like this, specially when problems arise.

	The actual code for the plugin is a Python regular package, and it can be as complex and needed, allowing for the use of sub modules, importing packages, installing them (via {\tt requirements.txt} file) and so on.

	They also offer support for other desirable capabilities such as sub commands, argument splitting, and regular expressions.

You can communicate with the bot sending text messages; the bot can also reply or send messages under some conditions. These replies  and messages from the chatbot support templating, using {\tt Jinja2}~\citep{jinja}.

	Finally, there is a persistence mechanism to store relevant information (the configuration of the plugin, for example, can be done online, and the actual data is stored in the local machine).

	There are other capabilities that we have not explored, such as {\tt streams}, which allow to be used for storing files (documents, generated content,..).

	Let us now see the actual code of some function to check for the CO2 concentration measured with one of our sensors. Since there are good libraries to manage the MH-Z19 in this case the code is deceptively simple, as it can be seen here.

\begin{lstlisting}[language=Python,firstline=36,lastline=42, caption={Code for reading the value of a sensor}%,keywordstyle=\color{black},
  %commentstyle=\color{black},
  %stringstyle=\color{black},
	]

    @botcmd(split_args_with=None) 
    def co2(self, message, args): 
        """A command which returns CO2 measurement""" 
        import mh_z19 
        res = mh_z19.read_all()
        return res
\end{lstlisting}


	In this case no special formatting is used, so we simply return the value received from the \lstinline+mh_z29.read_all()+ function.
	Neither no formatting or selection of information is done.

\subsection{Our Experiments\label{secResults}}

We have installed Errbot in several machines and configured it to interact in different channels (we have bots on IRC, Slack, Discord, Gitter, and Telegram):
being a proof of concept, our idea was to have a set of federated chatbots that can pass information among them and to have as many computers reporting information as needed. They are in different networks with diverse access control policies (some in public networks, some in private ones). The user can ignore all of these settings and concentrate on requesting and reading the information.

Of course, a more centralised approach is possible (just one channel, for example).
It  could also be possible to obtain the information (or a part of it) directly from some ERP (if available) installing the bot in some machine or maybe use a mixed approach.

Some plugins need to be developed in order to provide the needed information to the users.
The command  \& control part of the system, where chatbots write and read information needed to communicate was developed as an Errbot plugin for this project.

\subsection{Some Examples Of Interactions\label{subSecExamples}}

Given the fact that we want to hide as many difficulties as possible to the end user and that we can have a set of federated bots providing information, we need to establish two ways of accessing the information:
	\begin{itemize}
		\item The user contacts directly to the bot responsible of the relevant information.

		\item The user contacts some bot, which will forward the command and retrieve the information from the bot that can access it.
	\end{itemize}

In the first case, the user just needs to know the way to communicate with the chatbot (the messaging system and the identifier) writing the adequate command to request directly the information.

	For example, he/she could write the following previously shown command \lstinline+temp+ directly to the bot interface to get the temperature from some sensor:

	\begin{quote}
		\lstinline+temp+
	\end{quote}

	And the reply could be:

	\begin{quote}
		\lstinline{Temp: 69.8 F / 21.0 C    Humidity: 57.1}
	\end{quote}

	This is a granular approach, adequate when the user just needs to monitor some resource (or a limited number of resources).

	In the second case, when several computers are involved and when the user needs accessing several of them we can have a set of federated chatbots.

	Let us remark that we have simulated a quite heterogeneous situation with several bots, communicating via different backends in order to evaluate the feasibility of this approach. A realistic solutions would probably be more simple, maybe with just one backend.

	Since we are proposing a set of bots and it can be difficult and
	uninteresting to know about all of them, we need some way to
	coordinate communications: typically our user will interact with one
	of the chatbots (or maybe different bots depending on its context: mobile devices when outside, a desktop computer when in the office, \ldots).
	Then, this bot will forward the commands to the
	others requesting the needed information from them.

	In order to simplify the setup we decided to avoid complex mechanisms for registering bots and raise awareness; the objective was to introduce simple and robust ways for reaching the desired results.
	For this, we are using command and control techniques.
	These  techniques have been used mainly in the malware and botnet management. We are proposing here to use them for good.

	In this context, bots are responsible of registering themselves in some common channel (more complex setups are possible, of course) when they are activated; they can write and read there. This is the way we are proposing so they can send and receive commands to/from other bots.
	From the point of view of the user he or she can communicate with one of them, and it will forward orders to this channel and reading later the results.

	This C\&C channel can be located in a public infrastructure (with some restrictions or not about who can read and write there, depending of it is set as public or private), with memory or not (that is, new bots can have access to all the previous communications log or not) and so on.
	We have done our tests using a Slack channel~\citep{Slack}. We also tried Gitter~\citep{Gitter} to be reasonably sure that the approach was independent of the platform: some place where the chatbots can write and read their text messages is needed.

	We could have tried some MQTT~\citep{Light17} server, for example, but this use of public infrastructure in order to simplify everything is quite attractive to us.
	If the platform provides memory (Slack and Gitter, for example, allow reading part of all the past messages in a given channel, respectively) the bot will be able to respond to older requests; if the platform has not memory, it won't (in IRC, for example, the bot has not access to the past conversation history).
	In this sense, we have selected Slack and Gitter as possible backends to allow for our bots to check past commands, in order to be able to reply to then even when they were not present when they were issued.

	We have developed a plugin for Errbot, \lstinline{err-forward},
	which can perform the following operations:

	\begin{itemize}
		\item {\bf Reading}:
			The bots can read messages in the channel. There can be several types of messages.
			\begin{itemize}
				\item Messages of presence ({\tt Msg}). Each bot can read all the messages published in the channel to see which bots have been registered. The only important use is to forward some message to some or all of them.
					Each bot does not record nor store the presence of other bots, just looks for messages of past registering when needed.
				\item Messages with commands ({\tt Cmd}): a bot will check if  it can execute the command with the arguments provided; if it can, it will delete the command from the channel and execute it.
				\item Messages with replies ({\tt Rep}) from other bots: if  the reply is addressed to the bot it will read it, delete it from the channel, and send it to the user.
			\end{itemize}

		\item {\bf Writing}:
		The bots can write messages in the channel. There are several types of messages.
			\begin{itemize}
			\item Registering ({\tt Msg}): a bot can write when it starts working, to show presence (this presence is never updated: the bots use an asynchronous model to communicate and, in particular, they do not check if the receiver of a command is available or not for the shake of simplicity).
			\item Instructions for other bots ({\tt Cmd}): a bot can write a command in order to pass it to other bots.
			\item The results of some command ({\tt Rep}) they have read and executed previously.
			\end{itemize}
	\end{itemize}

	So, the typical workflow for reading a sensor could be to send some message to our bot (one of them, as stated before). In the example, checking for the reading of the CO2 sensor available in some room:

	\begin{quote}
		\lstinline{fw co2 room23}
	\end{quote}

Notice that the forward command is not strictly needed, since they could post the command in the C\&C channel when they are not able to execute it.

	Our bot can be the responsible for executing this order, and then execute it and return a value, or it could just pass the command to the others (at the current state we would explicitly send a forward command similar to {\tt fw co2 room23}, but this could be avoided).
	In the second case, it would write in the common channel something like (we have selected JSON as message container):

	\begin{quote}
\begin{lstlisting}
{"userName": "someUserName",
 "userHost": "someUserHost",
 "frm": "someUserIdentifier",
 "typ": "Cmd",
 "cmd": "co2",
 "args": "room23"}
\end{lstlisting}
\end{quote}

	Here, the {\tt userName} and {\tt userHost} are identifiers for the sender, {\tt frm} is a channel-based identifier, {\tt typ} defines the type of the command ({\tt Cmd} is an order, we will see later a reply, {\tt Rep}), {\tt cmd} is the actual command ({\tt co2}, in this case), and {\tt args} allow us to add some parameters to the command ({\tt room23} in this example).

	The other bots will be able to read this command and check whether they can execute it or not (basically replying to two the questions stated in the Section explaning the architecture. Notice that in this case the command ({\tt Cmd}) is empty as the type ({\tt typ}) is a reply ({\tt Rep}) and the actual content of the reply is part of the argument, ({\tt args}). Notice that this argument is encoded in order to allow any set of characters without problems in web and textual platforms.

		\begin{quote}
\begin{lstlisting}
{"userName": "otherUsername",
 "userHost": "otherHost",
 "frm": "someUserIdentifier",
 "typ": "Rep",
 "cmd": "",
 "args": "... co2 ... temperature%27...%7D"}
\end{lstlisting}

	\end{quote}

	In this case {\tt userName}, {\tt userHost} and {\tt frm} are as in the previous step,  adequate identifiers. Notice that the later is used to keep record of the original sender, in order to be able to read the reply.

	Finally, the first bot will perform the adequate actions to read, decode and return the reply to the user who made the original petition.


	In this case the output is not very elaborated and it just shows all the parameters provided by the sensor.

	\begin{quote}
\begin{lstlisting}
{'co2': 1099,
 'temperature': 26,
 'TT': 66,
 'SS': 0,
 'UhUl': 4608}
\end{lstlisting}
	\end{quote}

	The self-registering process is simple: the bot just writes a `Hello' message in the channel, with some context information.

	\begin{quote}
\begin{lstlisting}
{"userName": "", "userHost": "",
 "frm": "", "typ": "Msg", "cmd": "",
 "args": "Hello! IP: 192.168.1.82. Commands [/]. Name: estudio. Backend: Discord"}
 \end{lstlisting}
	\end{quote}

	Notice that in this case there is information about the backend used (Discord) and the IP address (a local one in this case).

	Related to this we can have also other command to see which bots are in the channel (based on when they registered), {\tt listB}, but it is more interesting for debugging matters and management than for the final user.

	Adding new commands, new computers, and/or new communication channels is straightforward and does not impose any modifications in the available infrastructure or the chatbots that are operating at any given moment. The changes are local to the involved computer, and the rest does not need to manage them.

\section{DISCUSSION\label{secDiscussion}}

Accessing sensor data or other resources across different networks and locations can be challenging. Our first choice could be setting a web page or an app with a dashboard, but this can be time consuming and even difficult to do in some occasions.

The proposed approach is quite simple to use, it has low coupling with other systems and provides a reasonable level of security.
Moreover, it can be very adequate for occasional use, where a quick setup is needed, with no need to change the information systems and in such a way that the information is accessible almost to anyone.

It can also be a sustainable long-term solution: the infrastructure needed is very simple and almost any computer can do the task. This does not mean that it is a limited solution, since  the chatbot could be integrated with other enterprise-level software, or databases without much problem.

It has some limitations, anyway:

\begin{itemize}
	\item We are proposing a text-based interface and this won't be adequate for everybody. With some messaging systems there could prepare buttons (Telegram, or Slack for example have them) but it also would add complexity and developing time.

	\item Commands (text-commands) must be well thought in order to be useful without being difficult to type (if the user is in a limited device, such a smartphone) and easy to remember (of course the user can see help information, but in our experience users tend to avoid it).
So they should be easy to type, short, meaningful...

		\end{itemize}

 \section{CONCLUSIONS\label{secConclusions}}

 An architecture proposal for federating chatbots has been presented.
  It includes the use of heterogeneous infrastructure, simple devices and it can be integrated in our standard architecture.

 For this the use of a set of federated chatbots using command and control techniques for controlling and reporting has been tested in a home/laboratory environment.
 The approach seems to be simple but powerful and it can provide an entry point for starting the integration for the access to several sources of information.
%
%
%

 \section{FUTURE WORK}

Future plans for this project are related to integrating more sources of information (sensors, but also internet sources, APIs, ...)
The goal is to have an information manager which can be accessed in an easy way with a simple device such as an smart phone without the burden of installing apps or visiting web pages (and with the advantage of not needing to change our access model inside the firewall).

Another possible idea would be to allow for more complex workflows: maybe one chatbot replies to some command, and there is another (or several ones) that can process this reply, or add more information, etc.

In this presentation we have concentrated in a proof of concept with some simple use cases, but a world of possibilities is available with this technology.

 \section{ACKNOWLEDGEMENTS}
The author is with the Computer Science for Complex System Modeling (COSMOS, T64\_23R), partially co-financed by the Aragonese Government and the European Social Fund.
He also is working in the DL-Ageing project under grant PID2019-104358RB-I00 from the Spanish `Ministerio de Ciencia e Innovación'.

\section*{AUTHOR BIOGRAPHY}

\noindent {\bf FERNANDO TRICAS} is an associate professor in the
Departamento de Informática e Ingeniería de Sistemas at Universidad de Zaragozaz.
His research interests include analysis and modeling of concurrent systems using formal methods.
His e-mail address is
\url{<ftricas@unizar.es>} and his Web address is
\url{<http://webdiis.unizar.es/~ftricas/>}.




%
%
%
%
\bibliographystyle{eurosis}
\bibliography{chatbots}
%
%
%
%
\end{document}